\documentclass[12pt,a4paper]{article}
\usepackage{graphicx}
\usepackage{times}
\title{\bf Hot colliding winds and the 2009 campaign on WR140}
\author{Anthony F. J. Moffat \\
\\
\normalsize D\'epartement de physique,  Universit\'e de Montr\'eal,  \\
\normalsize and Centre de Recherche en Astrophysique du Qu\'ebec, \\
\normalsize  C.P. 6128, Succ. Centre-Ville, Montr\'eal, QC, H3C 3J7, Canada
\\
\\
\normalsize Published in proceedings of \\
\normalsize"Stellar Winds in Interaction", editors T. Eversberg and J.H. Knapen. \\ 
\normalsize Full proceedings volume is available on http://www.stsci.de/pdf/arrabida.pdf
}

\date{\mbox{}}
\begin{document}
\maketitle
%\pagestyle{empty}
%
% WE REDEFINE THE plain LaTeX PAGESTYLE !!! 
% THIS PAGESTYLE WILL BE USED FOR THE FIRST PAGE ONLY !
%
\def\bull{\vrule height .9ex width .8ex depth -.1ex}
\makeatletter
\def\ps@plain{\let\@mkboth\gobbletwo
\def\@oddhead{}\def\@oddfoot{\hfil\tiny\bull\quad
Workshop ``Stellar Winds in Interaction'' Convento da Arr\'abida, 2010 May 29 - June 2\quad\bull}%
\def\@evenhead{}\let\@evenfoot\@oddfoot}
\makeatother
%
% AND DEFINE OUR MACROS FOR THE REFERENCE LIST
% I.E \beginrefer \refer and \endrefer
%
\def\beginrefer{\section*{References}%
\begin{quotation}\mbox{}\par}
\def\refer#1\par{{\setlength{\parindent}{-\leftmargin}\indent#1\par}}
\def\endrefer{\end{quotation}}
%
% BEGIN THE ABSTRACT CHAPTER WITH \noindent\small, ENCLOSE IT IN A GROUP
% AND BOLDFACE THE TITLE.
%
{\noindent\small{\bf Abstract:} 
WR140 (WC7pd + O5) is often considered to be the archetype of hot, luminous colliding-wind binaries, with strong cyclic high-energy and dust-formation events. The challenge is that this system is quite extreme, with a long period (nearly an integral 7.94 years) and high eccentricity (e = 0.88). Most of the action thus occurs during the relatively short several-month interval of close periastron passage, which in the most recent 2009 January passage occurred during the northern winter months when this summer Cygnus star was least favourably placed in the sky for groundbased observation.  To meet this challenge, various multiwavelength campaigns were organized at different sites and from space, mainly within several months of periastron passage 2009.  Of particular interest to this workshop was the MONS optical spectroscopic effort, involving both amateur and professional astronomers at the MONS site on Tenerife from 2008 December through 2009 March.  I will describe WR140 in terms of the general phenomenon of hot colliding winds and leave the new campaign details to other speakers at this workshop.
}
%
% NOW COMES THE MAIN BODY OF THE ARTICLE
%

\section{Preamble} 
Among high-energy phenomena in the Universe, colliding winds (CWs) occupy a prominent place.  While they are not in the highest-energy category (MeV and higher, e.g. gamma rays) associated with relatively rare events such as supernovae (SNe) and gamma-ray bursts (GRB), CWs produce energies in the several keV range (e.g. X-rays, corresponding to temperatures of several tens of million Kelvins) as stellar winds at velocities of order 1000 km/s collide head-on.  They are quite common, occurring in essentially all binaries (themselves very frequent) containing massive stars with strong, fast winds and in young regions of high stellar density, where multiple winds collide to produce hot intra-cluster gas. This article deals only with binaries.

\section{Massive stars}
What is meant by a ``massive" star? Normally one takes massive stars to be those stars with initial masses above 8 $M_{\rm Sun}$.  Such stars can, in their cores, nuclearly ``burn" H into He, He to C, etc., and finally ending up with Fe, at which point no more energy can be extracted from this process and they end their lives violently as core-collapse supernova (ccsn) (or even earlier as pair-instability supernova - pisn - for the most massive objects).  The former (ccsn) leave neutron stars (NS)or black holes (BH, for more massive progenitors), while the latter (pisn) blows the whole star apart.
What distinguishes massive from other stars, besides their hot temperatures (during most of their lifetime) and extremely high luminosities, is their ability to drive off matter in the form of strong stellar winds.  While the Sun has a prominent wind of particles with typical speeds of c. 500 km/s, its mass-loss rate is relatively low [c. $10^{-14}$ $M_{\rm Sun}$/yr].  Only massive stars can produce high-speed (up to 5000 km/s), dense outflows [up to $10^{-4}$ $M_{\rm Sun}$/yr], driven by the radiation pressure of their extreme luminosities.  Hence, despite their rarity, massive stars are the main drivers of the ecology of the Universe, via both stellar winds during their whole lifetimes with an ever richer mix of heavier elements the more the star evolves, and their SN explosions. Normally, the winds are strongest towards the end of their evolution, just before the SN explosion. This occurs during the core He-burning Wolf-Rayet (WR) stage (subsequent rapid core burning even up to Fe can also be occurring without being noticed at the stellar surface before it is too late), which lasts about 10\% of the star's whole lifetime.  For WR stars the mass-loss rates are typically at least an order of magnitude higher than their core H-burning O-star progenitors.  Within the WR stage, the WN phase reveals H-burning products (mainly He and N) in its wind, while the subsequent WC or WO phase reveals C- and O-rich winds from core He-burning.

\section{Massive binaries}
The majority (possibly even the entirety at birth) of massive stars are found in multiple systems, with binaries the most frequent and most stable configuration.  Among these massive binaries, one recognizes two grand classes (Vanbeveren, Van Rensbergen \& De Loore 1998):  (1) non-interacting, with initial periods $P \geq P_{\rm c} = 10$ years, and (2) interacting, with $P \leq P_{\rm c}$.  The former (1) behave essentially like two single stars, such that neither star in the binary expands enough in its evolution to spill over its Roche lobe overflow (RLOF) onto the companion.  The latter (2) can produce RLOF (with some complications, e.g. stars above initial mass $\sim$ 25 $M_{\rm Sun}$ never reach the large red supergiant stage, maxing out as much smaller LBVs, so Pc for them could be much smaller than 10 years), sooner for shorter initial periods, and even a common-envelope configuration for systems with the shortest periods and high mass ratios.  But all massive binary systems can, in principal, produce CWs, which is the subject of the rest of this article, although the impact diminishes roughly as the inverse separation of the two stars.

It is interesting to ponder what one would miss if it weren't for massive binaries, particularly of the interacting kind.  Following Moffat (2008) these include: enhanced stellar X-rays, non-thermal  radio emission, WR dust spirals, inverse mass-ratios, very rapid stellar spin, rejuvenation \& blue stragglers, enhanced dense-cluster dynamics, massive runaways, intermediate- \& supermassive-BHs,  gamma-ray bursts (GRB), and facilitating obtaining certain stellar parameters such as masses and sizes.  Clearly the gain in information far exceeds a factor two, i.e. merely going from one to two stars!

As an example of the first phenomenon, X-rays are produced by shock thermalization of plasma flows, with $kT \sim mp v^2 /2$ ($k$ is the Boltzmann constant, $T$ the temperature, $m_{\rm p}$ the proton mass, $v$ the velocity of the gas that is stopped).  This yields $T \sim  10^5 $K for $v \sim 100$ km/s (e.g. internal shocks in hot winds) and $T \sim 10^7$ K for $v \sim 1000$ km/s (e.g. CWs in hot, luminous binaries). As for GRB, all extremely rare although detectable throughout the Universe, those of long duration are believed to involve a SN of type Ic (i.e. lacking H and He) in a rapidly rotating WC/O star, possibly spun up in a very short-period binary. Short GRB likely occur during the rapid final merging of a NS+NS or NS+BH system, resulting ultimately from previously exploded massive stars.

\section{Colliding winds}
\begin{figure}[h]
\centering
\includegraphics[height=7cm]{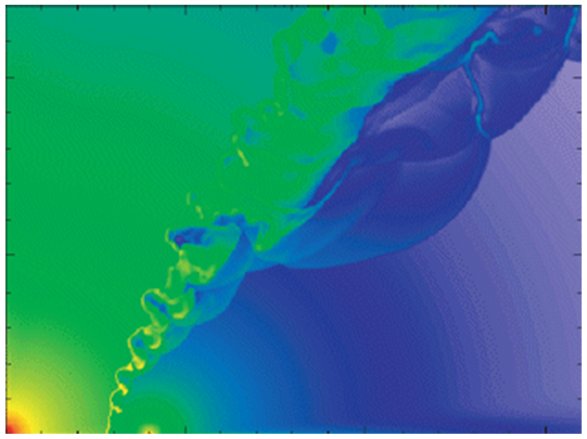}
\caption{\label{mof1}
Numerical simulation of a WR (centred at left corner) + O (centred at 4th tick to the right) CW system (Pittard priv. comm.). Density decreases from red, yellow, green to blue.}
\end{figure}

In the simple case of two massive, hot luminous stars with winds in a binary, it is inevitable that the two winds will collide, no matter what the separation .  For two spherically symmetric, uniform winds, the resulting collision will produce a shock cone defined primarily by a contact surface, where the rate of momentum transfer of each star's wind cancels vectorally.  On either side of this surface, there will be a shock front of hot plasma associated with each star.  The shape of the contact surface can be calculated algebraically (e.g. Usov 1992; Canto, Raga \& Wilkin 1996) or numerically (e.g. Stevens et al. 1992 - see Fig. \ref{mof1}); basically it resembles a hyperbolic cone of rotation, rounded off at the apex, with its axis in the orbital plane pointing towards the star with the stronger wind.  The cone opening angle depends in a predictable way on the ratio of the rate of momentum transfer of each star's wind, at the moment of contact (usually at terminal speed except in very close systems). The shape of the cone will change if the winds are not spherically symmetric, but we will not worry about that for the moment.  

The evidence for the existence of CWs has been somewhat slow in coming.  However, we now know that CWs produce excess X-rays of higher energy than the stellar winds (although sometimes difficult to detect against the stars' own X-ray emission), non-thermal radio emission (from relativistic acceleration of electrons in the CW shocked region in a locally produced magnetic field), and excess line emission in the UV/optical/NIR (and even X-ray lines!).   The strongest effects are seen in binaries containing at least one WR star, due to the very dense winds of these stars. The excess emission appears to fall off directly as the inverse of the orbital separation for large separations, where the interaction is adiabatic.  For closer systems, where radiative effects are more prominent, the fall-off with separation is faster.  Originally, X-ray and radio imaging was unresolved, but this is changing (to great benefit of constraining the geometry of the cones and the emission mechanism) with better telescopes (e.g. Chandra's 0.5\,arcsec resolution or VLBI imaging).  In the UV/optical/NIR, spectroscopy is not yet spatially resolved, although the spectroscopic Doppler effect effectively yields a kind of poor-man's resolution.

In some cases, one sees IR emission from warm carbon-based dust when the WR star is a carbon-rich WC. In the case of established binarity for many of these ``dustars", the dust must somehow be formed in the CW shock zone, where compression factors are high and shielding from the lethal dust-busting UV radiation occurs. But the exact process of dust formation remains a mystery, so any leads would be useful.  

\begin{figure}[h]
\centering
\includegraphics[height=7cm]{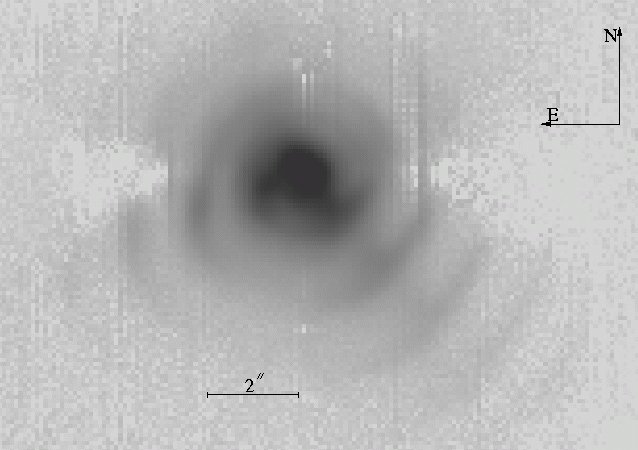}
\caption{\label{mof2}
WC9d star WR112 imaged in the MIR, showing 5 successive dust spirals, although broken several times in azimuthal distribution (Marchenko \& Moffat 2007).}
\end{figure}

In fact, there are two classes of WC stars that form C-based dust:  A. - WC+O binaries with WC of any subtype, although not all WC binaries do it.  B. - cool WC stars (usually WC9, but also a few WC8), whose binary nature is suspected in a number of cases, but not yet firmly established as a unique source in all cases.  Recent high spatial resolution work in the NIR/MIR  has revealed that many dustars are surrounded by resolved pinwheels, clearly the result of hot emitting dust flowing out and cooling down along the shock cone, which is wrapping up as the binary turns.  When the orbit is highly elliptic, as in the frequent case of long periods $\geq$ a year, the resulting spirals, although repeating faithfully from one turn to the next, appear broken in symmetry (see Fig. \ref{mof2}).  One way to explain this is via misaligned axisymmetric CWs, which would lead to four directions of enhance dust, much as one sees in certain systems (Moffat, in prep.).

\section{The case of WR 140}
\begin{figure}[h]
\centering
\includegraphics[height=7cm]{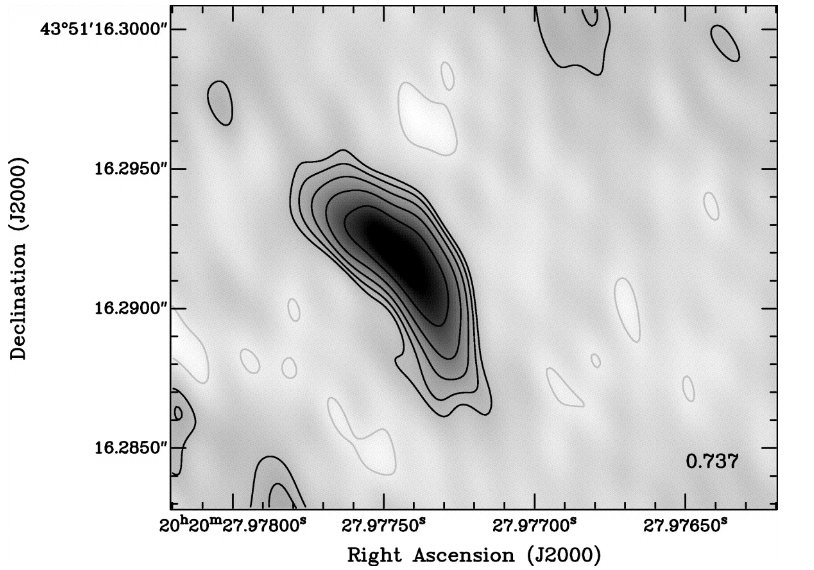}
\caption{\label{mof3}
VLBI imaging snapshot of WR140 of the resolved bowshock head in non-thermal radio emission (Dougherty et al. 2005).}
\end{figure}

\begin{figure}[h]
\centering
\includegraphics[height=7cm]{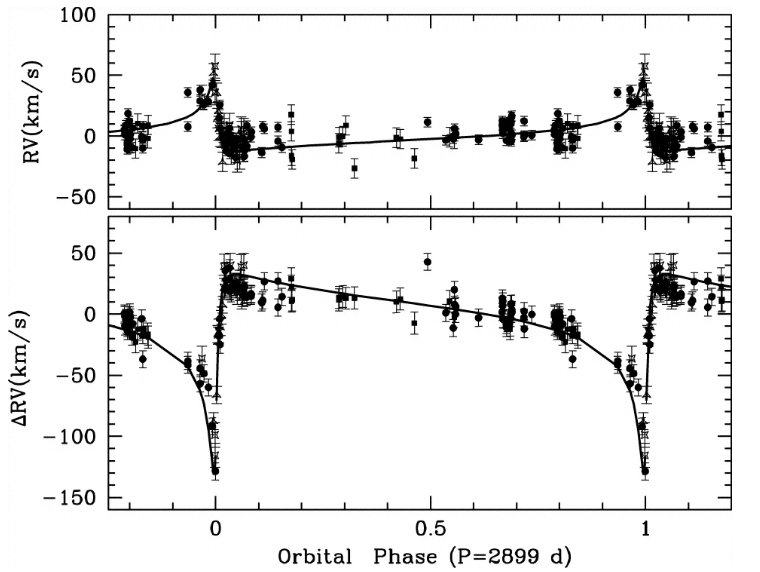}
\caption{\label{mof4}
RV orbits of the two components (O star above, WR star below) of WR140 from the 2001 campaign (Marchenko et al. 2003).}
\end{figure}

Located in the Cygnus direction of the Milky Way at a distance of just under 2 kpc, WR140 = HD193793, WC7pd + O5 (with ``p" referring to the WR star's peculiarly broad emission lines for its subtype and ``d" referring to its dust-making), is the brightest WR star in the northern sky, whose binarity defied detection for many years until finally in the 1980s its long ($P = 7.94$ year), highly elliptical ($e = 0.88$) orbit was discovered.  This allowed for a clear explanation of the NIR/MIR dust ``event" first discovered in 1977, coinciding with the time just after periastron passage when the winds collide with the greatest force. Indeed, the next IR events repeated in 1985 and 1993, so a multi-wavelength campaign was organized to study the nest 2001 periastron passage (where most of the action occurs) in great detail.  This included observational campaigns in the NIR, MIR, radio, X-ray and optical, all most intensely for several months on either side of periastron passage in mid February - a difficult time in mid winter to observe a summer object of modest declination (+40 degrees)! Of particular interest from that campaign and other recent data are: the high-resolution NIR/MIR images showing expanding arcs that repeat in the previous ejection (Williams et al. 2009); a resolved bow-shock head in non-thermal radio emission (see Fig. \ref{mof3}) coming up to, but dying away as self absorption occurs near periastron (Doughtery et al. 2005); a general $\sim1/D$ increase in X-ray flux towards and away from periastron, but with a narrow dip in X-ray flux just after periastron when the O-star eclipses the strongly X-ray emitting bow-shock head (Pollock et al. 2005); detection for the first time of optical emission-line excess in the density-sensitive line of CIII 5696A and the low-ionization line of HeI 5876A, giving constraints on the shock-cone properties, not to mention a vastly improved radial velocity (RV) orbit of both stars (see Fig. \ref{mof4}), along with mass-estimates of 50 and 19 $M_{\rm Sun}$ for the O and WR components, respectively (Marchenko et al. 2003). Optical broad-band monitoring also revealed several ~week-long, $\sim$0.1 mag deep dips over the 2-5 month interval after periastron; their colour dependence constrained the dust to be unexpectedly fine, with typical grain size 0.07 micrometres.

\section{Need for another WR140 campaign around periastron passage}
While the 2001 campaign was a success, we were still working somewhat blindly and it did raise several questions.  Most of all, we were amazed at how short the interval of optical emission-line excess was, leading to only a few points on the critical interval of $\sim$2 months on either side of periastron.  More intense coverage would be necessary in the next (2009) campaign to provide better constraints.  It would also be useful to know if the dust dips in the optical light curve 2-5 months after periastron repeat exactly as before or are a stochastic phenomenon.  We also needed denser IR coverage to constrain the shock-cone geometry, without which it will be difficult to trace the dust formation.  Perhaps WR140 will eventually solve this infamous problem, given its high eccentricity and thus ability to probe CWs at a vast range of separations in the same system.  Denser coverage is also required in X-ray imagery and spectroscopy and in radio imaging at multi frequencies.

Due to the near integer period in years of WR140, the 2009 periastron passage in mid January provided the same challenges as that of 2001.  Basically, from the ground one will only be able to observe WR140 for a small portion of the night, the more periastron is approached  from either side (evening just after sunset or morning just before sunrise). Therefore, to fill the rest of the night in the MONS campaign, we monitored other luminous hot stars that are known to vary on timescales (days-weeks) unique to such a campaign.  These will be reported in later publications.

\section{Conclusions}
WR140 shows clear, intense CW action at all wavelengths during a short interval of its long orbit, centred on or slightly after periastron passage.  Other articles in this series following this one will discuss the results found so far in the 2009 campaign.

%
% USE A SECTION WITHOUT NUMBER FOR THE ACKNOWLEDGEMENTS
%
\footnotesize
\beginrefer

\refer Canto, J., Raga, A.C., Wilkin, F.P. 1996, ApJ, 469, 729

\refer Dougherty, S.M., Beasley, A.J., Claussen, M.J., Zauderer, B.A., Bolingbroke, N.J.  2005, ApJ, 623, 447 

\refer Marchenko, SV,  Moffat, A.F.J., et al. 2003, ApJ, 596, 1295

\refer Marchenko,  S.V., Moffat, A.F.J.  2007, in Massive Stars in Interactive Binaries, ASPC 367, p.213

\refer Moffat, A.F.J. 2008, in Massive Stars as Cosmic Engines, proc. IAU Symp 250, p. 119 

\refer Pollock, A.M.T., Corcoran, M.F., Stevens, I.R., Williams, P.M. 2005, ApJ, 629, 482

\refer Stevens, I.R., Blondin, J.M., Pollock, A.M.T.   1992, ApJ, 386, 265 

\refer Usov, V.V. 1992, ApJ, 389,635

\refer Vanbeveren, D., Van Rensbergen W., De Loore, C. 1998, The Brightest Binaries, ApSpScRev, 232

\refer Williams, P.M., Marchenko, S.V., Marston, A.P., Moffat, A.F.J., Varricatt, W.P., Dougherty, S.M., Kidger, M.R., Morbidelli, L., Tapia, M. 2009, MNRAS, 395, 1749

\endrefer

\end{document}